\documentclass[sigplan,nonacm]{acmart}
\AtBeginDocument{%
  }

\usepackage{xspace}
\usepackage{multirow}
\usepackage[most]{tcolorbox}
\tcbuselibrary{theorems}
\tcbset{highlight math/.append style={left=0mm,right=0mm,top=-1mm,bottom=-1mm, colframe=white}}

\definecolor{myblue}{RGB}{222,236,254}

\begin{document}

\title{Position: Intelligent Coding Systems Should Write Programs with Justifications}

\author{Xiangzhe Xu}
\email{xu1415@purdue.edu}
\affiliation{%
  \institution{Purdue University}
  \country{USA}
}
\authornote{Both authors contributed equally.}

\author{Shiwei Feng}
\email{feng292@purdue.edu}
\affiliation{%
  \institution{Purdue University}
  \country{USA}
}
\authornotemark[1]

\author{Zian Su}
\email{su284@purdue.edu}
\affiliation{%
  \institution{Purdue University}
  \country{USA}
}

\author{Chengpeng Wang}
\email{wang6590@purdue.edu}
\affiliation{%
  \institution{Purdue University}
  \country{USA}
}

\author{Xiangyu Zhang}
\email{xyzhang@cs.purdue.edu}
\affiliation{%
  \institution{Purdue University}
  \country{USA}
}

\begin{abstract}
Intelligent coding systems are transforming software development by enabling users to specify code behavior in natural language. However, the opaque decision-making of AI-driven coders raises trust and usability concerns, particularly for non-expert users who cannot inspect low-level implementations. We argue that these systems should not only generate code but also produce clear, consistent justifications that bridge model reasoning and user understanding. To this end, we identify two critical justification properties—cognitive alignment and semantic faithfulness—and highlight the limitations of existing methods, including formal verification, static analysis, and post-hoc explainability. We advocate exploring neuro-symbolic approaches for justification generation, where symbolic constraints guide model behavior during training and program semantics are enriched through neural representations, enabling automated consistency checks at inference time.
\end{abstract}

\maketitle

\section{Introduction}

We argue that intelligent coding systems should produce code accompanied by clear justifications.

\begin{figure}
    \centering
    \includegraphics[width=0.97\linewidth]{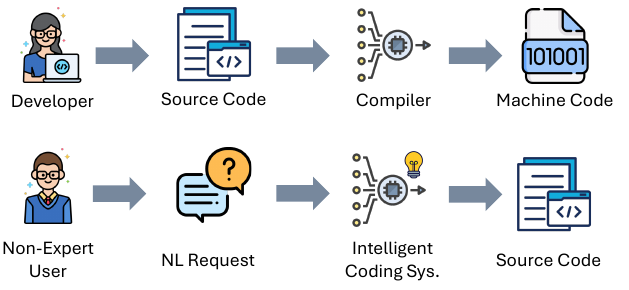}  
    \caption{Intelligent coding systems play a similar role as the traditional compilers. Traditional compilers empower developers by converting higher-level source code to executable machine code. Intelligent systems empower non-expert users by converting natural language requests to source code.}
    \label{fig:intro-compiler}
\end{figure}

By lowering the barrier to expert-level programming, intelligent coding systems democratize software development, enabling anyone to create complex code artifacts. These AI-driven tools can generate code from natural language descriptions~\cite{achiam2023gpt,claude,roziere2023code,abdin2024phi,team2024codegemma,xu2024prosec,wei2023magicoder,zhang2024seccoder,ding2024semcoder}, address bugs based on user-reported symptoms~\cite{xia2024agentless,yang2024swe,ma2024lingma,wang2024openhands}, and reason about program behavior from textual specifications~\cite{zheng2025validating,wang2024llmdfa,wang2024sanitizing,wang2024llmsa,guo2025repoaudit,pei2023can}. They translate diverse, high-level language requests into precise, executable code, bridging the gap between human intent and machine instructions. As illustrated in Figure~\ref{fig:intro-compiler}, in the same way that a compiler translates high-level programming languages into machine code, an intelligent coding system translates natural language prompts into precise programming code.

Programmers routinely write and execute test cases to verify that compiled programs behave as expected~\cite{ammann2017introduction,myers2011art,zhang2017skeletal}. Similarly, AI-driven coding systems require human supervision in critical scenarios: their probabilistic models can yield variable outputs for the same input~\cite{ouyang2025empirical,song2024good} and may reflect biases in their training data~\cite{xu2023improving,xu2023leveraging}. Consequently, human validation of generated code artifacts remains essential.

We believe that, in addition to producing code, an intelligent coding system should also generate \textit{justifications}.
Justification aims to provide clear, consistent explanations of AI-driven code generation, enabling users to understand and trust the produced artifacts by bridging the gap between opaque model reasoning and user comprehension.

A simple approach is to ask language models to generate their own `chain-of-thought'~\cite{wei2022chain} reasoning traces. These traces externalize the model’s internal logic, allowing humans to verify it~\cite{baker2025monitoring}. However, language models can produce unfaithful explanations that do not match their actual behavior~\cite{arcuschin2025chain,chen2025reasoning}, so code quality may still suffer even if the reasoning reads well.

We identify two essential properties for effective justifications:
(1) {\em Cognitive alignment}: Justifications should be expressed in natural language that aligns with human reasoning patterns, allowing non-expert users to follow and evaluate the system’s decisions without inspecting low-level code.
(2) {\em Semantic faithfulness}: Justifications must accurately reflect the system's internal process and be automatically verifiable against generated code entities, ensuring consistency between the explanation and the actual program semantics.

A promising direction is a neuro-symbolic design that integrates explainability~\cite{treviso2020explanation,lamm2021qed,dhaini2024explainability} techniques from the AI community with formal program semantics from the programming languages community~\cite{plotkin1981structural,pratt1976semantical,pierce2002types}. In this approach, semantic specifications of code serve as constraints and guidance during model training, helping the system generate justifications that meet domain requirements. Associating explanations with specific code entities then enables automated verification, ensuring the justifications are semantically faithful to the generated artifacts.

\section{Existing Efforts on Justification}

\begin{figure*}[t]
    \centering
    \includegraphics[width=0.8\linewidth]{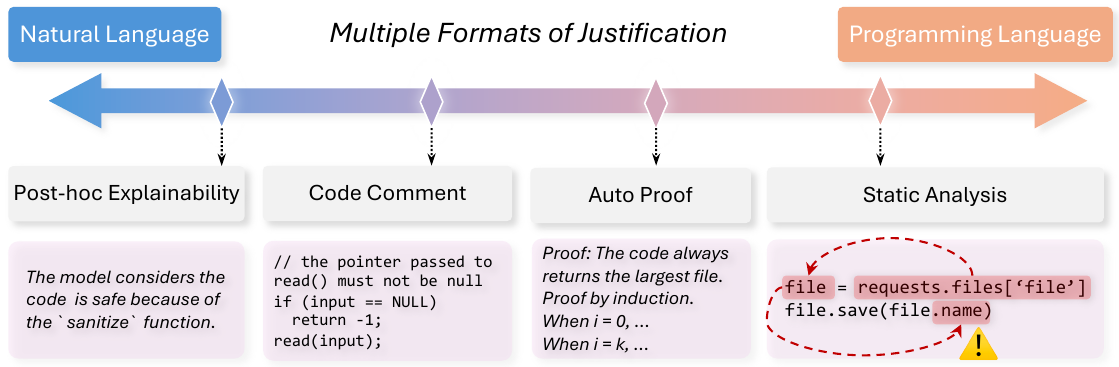}
    \caption{Existing efforts on generating justification for code artifacts. A technique closer to the left denotes the technique works closer to the natural language space, and vice versa, a technique closer to the right denotes it works at the level closer to the programming language space.}
    \label{fig:existing}
\end{figure*}

\begin{figure*}[t]
    \centering
    \includegraphics[width=0.8\linewidth]{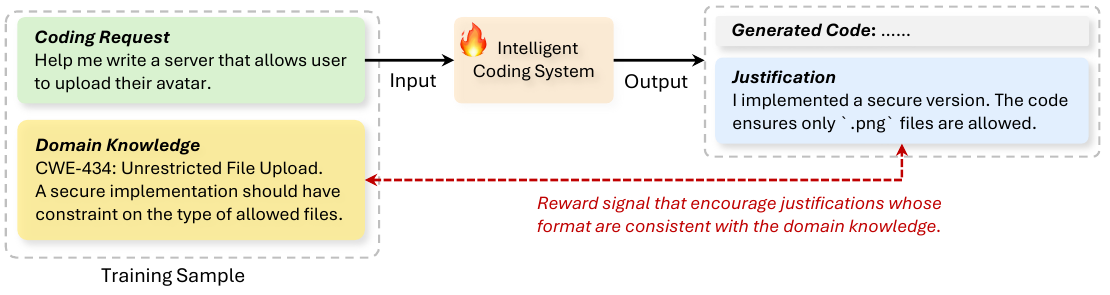}
    \caption{Reward signal for domain-aligned justification. For example, given the domain knowledge that avatar uploads should be restricted to \texttt{.png} files~(CWE-434), a coding agent tasked with building a user-avatar upload server must enforce file extension validation. The reward function grants positive feedback only if the justification explicitly discusses the \texttt{.png}-only constraint, ensuring consistency with domain knowledge.}
    \label{fig:our-tech-train0}
\end{figure*}

\begin{figure*}[t]
    \centering
    \includegraphics[width=0.8\linewidth]{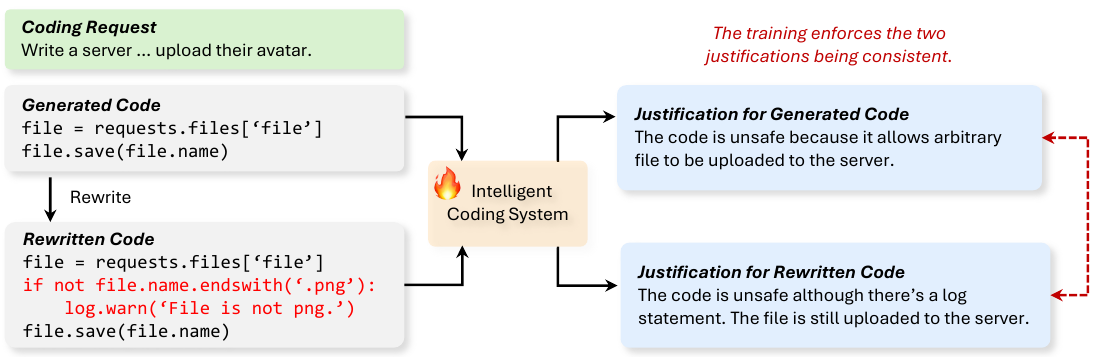}
    \caption{Metamorphic rewarding for justification consistency. By applying semantic-equivalent rewriting to a code snippet (e.g., one without any file-extension check and another that only logs a warning), our metamorphic rewarding mechanism guides the two corresponding justifications to be consistent.}
    \label{fig:our-tech-train1}
\end{figure*}

\begin{figure*}
    \centering
    \includegraphics[width=0.8\linewidth]{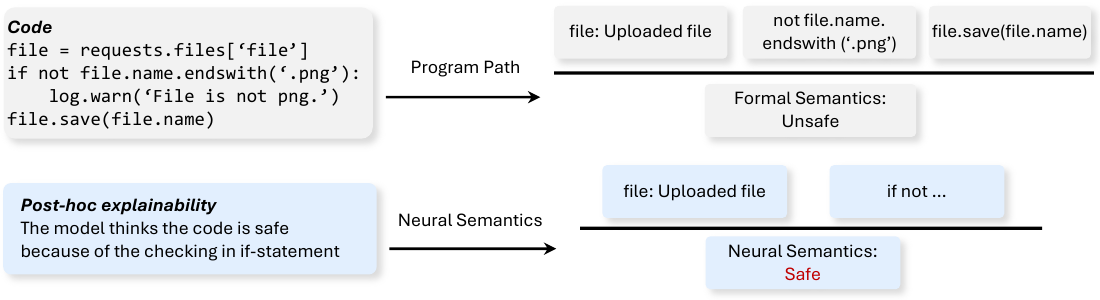}
    \caption{Neural-symbolic consistency check. The gray boxes describes a program path in which an uploaded file's name fails the {\tt .png} check yet is still saved, which is an unsafe behavior under formal semantics. In contrast, suppose that a post-hoc explainability indicates the model focuses on the presence of the if statement and labels the code as safe, as shown in the blue boxes. This discrepancy reveals that the model's justification is not semantically faithful to the actual program behavior.}
    \label{fig:our-tech-infer}
\end{figure*}

Generating justification for code artifacts is not a new concept in the programming language community. We show representative examples of justifications in Figure~\ref{fig:existing}. Static analyzers~\cite{moller2012static,wang2024llmdfa,wu2024libalchemy,tang2024octopus,tan2023tai,tan2022tai,shi2018pinpoint,sawant2022learning,bennett2024semgrep,pavela2023advanced} can quickly check whether a given code snippet meets specific requirements. For example, in Figure~\ref{fig:existing}, a static analyzer is used to check whether the name of an uploaded file is directly used to save the file without sufficient checks. Static analyzers are good at reasoning about programs at the level of programming languages (e.g., detecting buggy patterns), yet the higher-level properties (e.g., functional requirements~\cite{zheng2025validating}) may hardly be expressed or detected by a static analyzer.

Automated formal verification~\cite{xu2021automatic,fakhoury20243dgen,delaware2019narcissus,sanchez2023passport,heras2014ml4pg,kozyrev2024coqpilot,lu2024proof} can describe higher-level program semantics, yet it requires human expertise to write specifications in languages close to the programming language and use proof assistants to rigorously prove that a program satisfies certain properties. Such approaches demand substantial human effort: writing formal specifications and constructing proofs, which can be difficult to scale to non-expert users.

A justification that is closer to natural language is to use code comments. There are existing works that check the consistency between a natural language comment and the corresponding code snippet~\cite{tan2007icomment,zhai2020cpc,zhang2024leveraging,zhou2017analyzing,zhong2013detecting,stulova2020towards,rabbi2020detecting}. However, they typically focus on lower-level properties~(e.g., whether a pointer could be null or not; whether the usage of a given API is correct) rather than higher-level semantics.

Post-hoc explainability techniques for ML models aim to reveal how inputs influence outputs~\cite{lamm2021qed,yang2025prompts,errica2024did,zhuo2024prosa,seth2025xai_evals,lin2019explanations,salih2025perspective,sydorova2019interpretable,samadi2024safe,zhao2024explainability}. Representative techniques include gradient or attention-based indicators~\cite{salih2025perspective,gurrapu2023rationalization,bhan2023evaluating}, input perturbations (e.g., masking)~\cite{treviso2020explanation}, and training decoders on internal states~\cite{chen2025externalmonitorsenhancingtransparency}. Moreover, recent studies on chain-of-thought faithfulness~\cite{arcuschin2025chain} measure the alignment between generated reasoning traces and model outputs. However, these evaluations are limited to simple tasks, and extending them to complex coding artifacts remains an open challenge.

\section{Our Recommendations}

An effective justification should be accessible to non-expert users and remain consistent with the generated code artifacts. We recommend adopting a neuro-symbolic design: leveraging programming language domain knowledge to guide and constrain the training of intelligent coding systems, enabling them to generate plausible justifications that satisfy domain requirements. At the same time, linking justifications to code entities allows these explanations to be formalized and verified against rigorous program semantics.

\noindent
{\bf Training Guidance.} Large language models for code commonly use reinforcement learning to align model outputs with desired behaviors~\cite{ouyang2022training}. For each problem instance, the model samples multiple candidate solutions, evaluates each via a reward function~\cite{liu2024learning} that captures both code correctness and explanation quality, and then updates its parameters to favor higher-reward responses.

In domain-specific scenarios~(such as generating justifications alongside code), one key challenge is designing a reward function~\cite{liu2024learning,xu2024prosec} that optimizes for cognitive alignment and semantic faithfulness. We outline two approaches to constructing such reward functions~(see Figures~\ref{fig:our-tech-train0}~and~\ref{fig:our-tech-train1}).

\noindent
{\bf Inference-Time Verification.} To ensure justifications remain consistent with generated artifacts at inference, we propose a neural-symbolic verification pipeline. First, post-hoc explainability techniques (e.g., attention attribution or counterfactual perturbations) associate each element of the natural language justification with specific program entities in the code snippets. Next, we express the model’s reasoning as neural semantic rules—akin to inference rules in programming languages—whose premises reference those same entities. Finally, we compare these rules against the program’s formal semantics to detect any discrepancies. Figure~\ref{fig:our-tech-infer} provides a concrete example of this process.

\bibliographystyle{ACM-Reference-Format}
\bibliography{sample-base}

\appendix

\end{document}